\documentclass[aps,showpacs,prl,amsmath,amssymb,twocolumn,superscriptaddress]{revtex4}
\usepackage{graphicx}
\usepackage{subfigure}

\begin{document}
\title{Quantum Hall Superfluids in Topological Insulator Thin Films}
\author{Dagim Tilahun}
\affiliation{Department of Physics, Texas State University, San Marcos, TX 78666}
\affiliation{Department of Physics, The University of Texas at Austin, Austin, TX 78712}
\author{Byounghak Lee} 
\affiliation{Department of Physics, Texas State University, San Marcos, TX 78666}
\affiliation{Department of Physics, The University of Texas at Austin, Austin, TX 78712}
\author{E. M. Hankiewicz} 
\affiliation{Institut f\"{u}r Theoretische Physik und Astrophysik,
Universit\"{a}t W\"{u}rzburg, Am Hubland, 97074 W\"{u}rzburg, Germany} 
\author{A. H. MacDonald}
\affiliation{Department of Physics, The University of Texas at Austin, Austin, TX 78712}

\begin{abstract}
Three-dimensional topological insulators have protected Dirac-cone surface states.
In this paper we propose magnetic field induced topological insulator thin film ordered states in which coherence
is established spontaneously between top and bottom surfaces.  We find that the large dielectric constants of these materials increases the layer separation range over which coherence survives and decreases the superfluid sound velocity, but has little influence on superfluid density or charge gap.  The coherent state at 
total Landau-level filling factor $\nu_T=0$ is predicted to be free of edge modes, qualitatively altering its transport phenomenology. 

\end{abstract}
\maketitle

\noindent
{\em Introduction}---Double well two-dimensional electron systems in which Landau levels that are localized in different layers approach degeneracy tend \cite{PhysRevB.40.1087, PhysRevB.63.035305} to form broken symmetry states with spontaneous interlayer coherence \cite{PhysRevLett.69.1811} when one level is occupied.
These states support counterflow superfluidity and are 
sometimes referred to as quantum Hall superfluids \cite{MacDonald2001129}, although they can also be viewed as exciton condensates or as pseudospin ferromagnets \cite{PhysRevB.40.1087, PhysRevLett.65.775, PhysRevB.42.3224, PhysRevB.51.5138, PhysRevB.63.035305, eisenstein:2004}.  Their order gives rise to a number of subtle and interesting anomalies
in transport experiments \cite{ PhysRevLett.72.728, PhysRevLett.84.5808, PhysRevLett.91.076802, PhysRevLett.93.036802, PhysRevLett.93.266805, PhysRevB.70.081303, PhysRevB.77.033306, PhysRevB.78.075302, PhysRevB.78.205310, PhysRevB.80.165120} in which the two layers
are contacted separately that have motivated considerable theoretical activity \cite{PhysRevB.51.11152, PhysRevB.54.11644, PhysRevLett.86.1825, PhysRevLett.86.1829, PhysRevLett.86.1833, PhysRevLett.87.196802, PhysRevLett.88.106801, PhysRevB.66.115320, PhysRevB.68.155318, PhysRevLett.93.126803, PhysRevLett.95.156802, PhysRevLett.95.266804, Khomeriki2006, fil:780, su:2008, PhysRevLett.101.046804,  PhysRevB.81.195314, dima, rosenow}. 
In this Letter we predict that quantum Hall superfluid states will also occur in topological insulator (TI) thin films and 
highlight important differences between the double well and TI-thin-film cases.   

In TI thin films the spontaneous coherence occurs between Landau levels localized on top and bottom surfaces.
The quantum Hall superfluid state \cite{PhysRevLett.72.728, PhysRevB.70.081303, PhysRevB.78.205310, PhysRevLett.100.106803, PhysRevLett.104.016801} appears at two dimensional (2D) layer separations $d$ less than $ \sim 2 \ell$
where $\ell \sim 25 {\rm nm} \, (B[{\rm Tesla}])^{-1/2}$ is the magnetic length.  It follows that TI 
quantum Hall superfluid behavior is a possibility only 
in samples thinner than about $20 {\rm nm}$, and therefore only in materials 
prepared by a thin film growth technique. 
Since topological insulators  \cite{qi:33,RevModPhys.82.3045} have protected surface states, 
spatially separated Landau levels are always 
present when the Fermi level lies in the bulk gap.
Moreover, because the surfaces have massless Dirac electronic structure, 
their $N=0$ Dirac-point Landau levels will be 
widely separated from other levels even at weak magnetic fields, making it easier to overcome disorder and reach the quantum Hall regime. 
Importantly quantum Hall superfluids in TI thin films,
unlike those studied in quantum wells, can occur at total Landau level filling factor $\nu_T=0$, thus sidestepping the dissipationless chiral-edge-state transport channel which can obscure \cite{su:2008} counterflow superfluidity in the $\nu_T \ne 0$ case.

Most properties of ideal quantum Hall superfluids are 
determined by a small set of parameters: the superfluid density $\rho$, the charged quasiparticle energy gap $\Delta$, the 
phonon velocity $v$, and the single-particle tunneling gap $\Delta_{SAS}$. $\Delta_{SAS}$ depends exponentially on the thickness of the film and is also 
sensitive to the relative position of the Fermi level on the two surfaces, whereas 
$\rho$, $\Delta$, and $v$ depend only on the electron-electron interaction strength and 
are less sensitve to film thickness.  Known topological insulators tend to have small bulk energy gaps, 
and therefore large dielectric constants, a property that is likely to persist as new materials are found,
making dielectric screening an important issue.  
We find that although the phonon velocity of a TI quantum Hall superfluid is strongly suppressed by 
dielectric screening, the superfluid density and the energy gap are weakly 
altered.  In fact we estimate that the Kosterlitz-Thouless transition temperatures for these 
superfluids can be larger than in semiconductor quantum well systems.  
In the following we explain the basis for these estimates,
point out some experimentally relevant differences between TI and 
semiconductor bilayer quantum Hall superfluids, and 
discuss the relationship between quantum Hall superfluids and zero-magnetic-field exciton condensates.

\noindent 
{\em TI-Thin-Film Quantum Hall Superfluid Energy Scales}---When Landau level mixing is neglected, 
the interaction dependent parameters can be estimated from the following 
mean-field-theory expressions \cite{PhysRevB.51.5138}: 
\begin{eqnarray} \label{params}
\rho &=& \frac{\ell^2}{32 \pi^2} \int dq \, q^3 \, V_{tb}(q) F_{tb}(q)  \, \exp(-q^2 \ell^2/2), \nonumber \\
\Delta &=& \frac{1}{2\pi} \int dq \, q \, V_{tb}(q) F_{tb}(q)  \, \exp(-q^2 \ell^2/2), \nonumber \\ 
v &=& \frac{1}{2\hbar} \sqrt{\frac{\rho \, e^2}{c}},
\end{eqnarray} 
where
\begin{equation} 
c =  \Big[ \frac{4 \ell^2}{e^2} \int dq \, q \big[V_{\sigma}(0)-V_{\sigma}(q)  \big ]\,  \exp(-q^2 \ell^2/2) \Big]^{-1}
\end{equation} 
is the bilayer capacitance per unit area,
and $V_{\sigma}(q) = (V_{tt}(q) F_{tt}(q) + V_{bb}(q) F_{bb}(q)  - 2 V_{tb}(q) F_{tb}(q) )/4$. In these equations $V_{tt}$,
$V_{bb}$ and $V_{tb}$ are respectively the interactions between two top surface electrons, two 
bottom surface electrons, and a top and a bottom surface electron.
$V_{\sigma}$ specifies the potential difference between surfaces generated by a charge 
transfer and the $F_{ab}(q)$ are form factors \cite{PhysRevLett.96.256602} which capture the influence 
of Landau-index dependent changes in single-particle wavefunction shape.  


\begin{figure}
\includegraphics[trim = 0.35in 2.25in 0.25in 2.25in, clip,width=3in,height=3in]{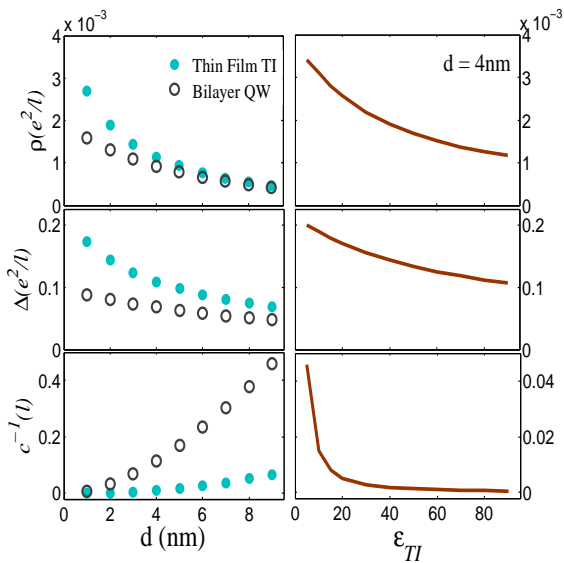}
\caption{Left panel: Superfluid density $\rho$, charge gap $\Delta$, and the 
reciprocal of the capacitance per unit area $c^{-1}$ as a function of TI thickness $d$. The solid dots were calculated with typical topological insulator parameters 
($\epsilon_{TI}=40$, $\epsilon_S=10$, and $B=10\rm{T}$) and the 
empty dots with GaAs double quantum well parameters.  
$\rho$ and $\Delta$ are larger than in the double quantum well case  in spite of the 
large value of $\epsilon_{TI}$.  The large TI dielectric constant enters however directly in the 
capacitance,  and therefore in the superfluid sound velocity.
Right panel: $\rho$, $\Delta$ and $c^{-1}$ as functions of $\epsilon_{TI}$ for $d=4\rm{nm}$. $\rho$ and $\Delta$ are much more weakly dependent on 
$\epsilon_{TI}$ than is $c^{-1}$.  The results in the right panel refer to the right vertical axis.}
\label{fig:rhodltcinv}
\end{figure}

It turns out to be very important for TI properties to understand how 
the values of these characteristic parameters are influenced by the TI bulk dielectric-constant ($\epsilon_{TI}$)
which is expected to be large, and also 
by the substrate dielectric constant $\epsilon_{S}$  on which the film sits.
An elementary calculation \cite{PhysRevB.82.085443} yields the screened potentials (assuming the top surface is exposed to vacuum)
\begin{equation} 
V_{tt}(q) = \frac{4 \pi e^2}{q D(q) } \; \big[ (\epsilon_{TI} + \epsilon_{S})\,e^{qd} + (\epsilon_{TI} - \epsilon_{S})\
 e^{-qd} \big],
\end{equation} 
\begin{equation} 
V_{tb}(q) = \frac{8 \pi e^2}{q D(q) } \;  \epsilon_{TI},
\end{equation}
and
\begin{equation} 
V_{bb}(q) = \frac{4 \pi e^2}{q D(q) } \; \big[ (\epsilon_{TI} - 1 )\,e^{qd} + (\epsilon_{TI} + 1)\
 e^{-qd} \big],
\end{equation} 
where $D(q) =  (\epsilon_{TI} + \epsilon_{S})  (\epsilon_{TI} +1 )\, e^{qd}+
(\epsilon_{TI} - \epsilon_{S})  (1-\epsilon_{TI} )\, e^{-qd}$.
When the film thickness coherence criterion is satisfied,
$qd$ will generally be small when the integrands in Eqs.(~\ref{params}) are large.
Setting $qd \to 0$ we find that $V_{tb}$, which appears in the expressions
for the superfluid density and charge gap goes to $4 \pi e^2/q(\epsilon_{S}+1)$, unaffected by $\epsilon_{TI}$. For $rho$ and $\Delta$, this leads to a weak dependence on $\epsilon_{TI}$ shown in the right panel of Fig.~\ref{fig:rhodltcinv}. In semiconductor quantum wells, $V_{tb} \to 2 \pi e^2/q\epsilon_{S}$, and therefore will tend to be, if anything, larger
for TI thin films than for typical double quantum wells ($\epsilon_S\approx 10$).  On the other hand 
$V_{\sigma} \to \pi e^2 d/\epsilon_{TI}$, which will tend to be strongly reduced in TI's 
compared to quantum wells, resulting in the suppression of $c^{-1}$ (Fig.~\ref{fig:rhodltcinv}, lowest panel), and subsequently the phonon velocity $v$.
In Fig. \ref{fig:rhodltcinv}, we plot $\rho$, $\Delta$, and the inverse capacitance $c^{-1}$ for some sample thicknesses and dielectric constants.

\begin{figure}
\includegraphics[trim = 1.0in 2.25in 1.0in 2.25in, width=3.0in,height=3.0in]{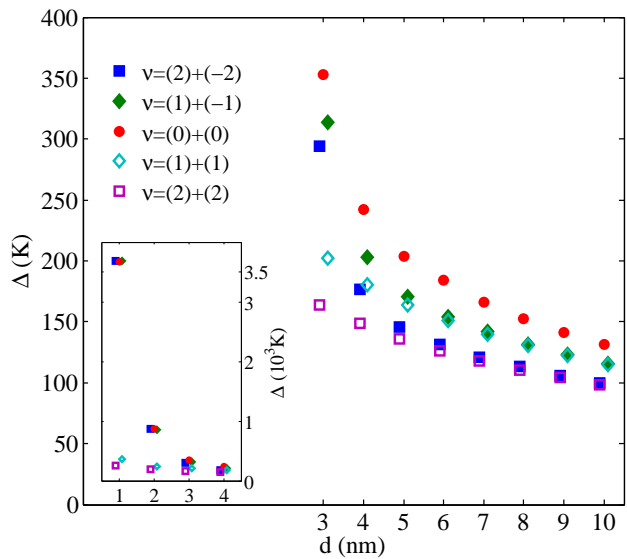}
\caption{Charge gap in Kelvin units as function of surface separation (TI thickness) for degenerate half-filled Landau levels. The gap values are larger than in the  
corresponding GaAs double-quantum well case.  These results were obtained with dielectric constant 
parameters $\epsilon_{TI}=40$ and $\epsilon_S=10$ and magnetic field $B=10\rm{T}$.
The labels specify the filling factor contributions $\nu_t,\nu_b$ from top and bottom surfaces
(or equivalently the orbital label of the partially filled level).
Inset: For samples thinner than $3\rm{nm}$ single-particle tunneling grows rapidly and dominates the gap
when $\nu_{t}=-\nu_{b}$.  Note that the energy scale for the inset is larger and specified on the 
right axis.} \label{fig:Nu024Gaps}
\end{figure}

\begin{figure}
\includegraphics[trim = 1.0in 2.25in 1.0in 2.25in, width=3in,height=3in]{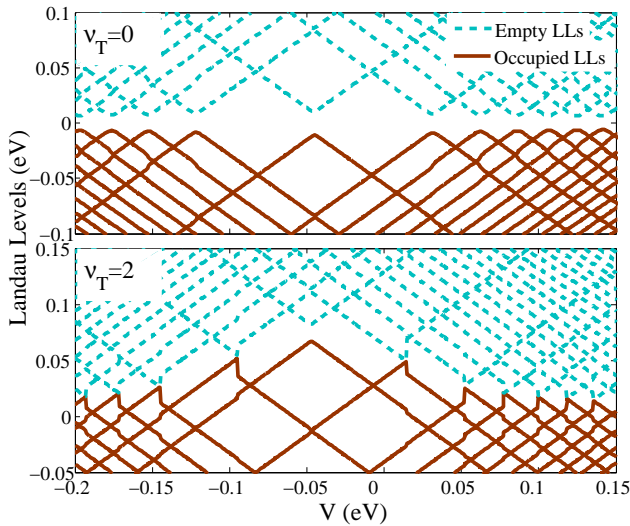}
\caption{Coulomb interaction modified Landau Levels for a TI thin film ($\epsilon_{TI}=40$ and $\epsilon_S=10$) of thickness $d=4\rm{nm}$ at $B=10\rm{T}$ as 
a function of external potential $V$.  Gaps open at the Fermi level due to interaction effects at
crossings between top and bottom layer Landau levels.   At this thickness, single-particle tunneling is small
compared to interaction induced gaps. In the top panel, corresponding to total filling factor $\nu_T=0$, gaps associated with inter-surface coherence occur over a broad range of $V$ values.  
In the bottom panel $\nu_{t} \ne - \nu_{b}$ so that the crossing Landau levels have different form
factors in all but the case when $\nu_{t} = \nu_{b}$. For this reason, the $\nu_{t} \ne \nu_{b}$ crossings gaps are associated with 
Ising pseudospin order involving jumps in the charge distribution across layers.} \label{fig:HFLLsvsV}
\end{figure}


\noindent
{\em TI Quantum Hall Superfluid Spectra}---In Fig.~\ref{fig:Nu024Gaps} we plot the film thickness dependence of the charge gaps that appear in TI thin films when two half-filled Landau levels are degenerate.  
The gaps plotted in Fig.~\ref{fig:Nu024Gaps}, evaluated including Landau level mixing \cite{blgevolution},
are in most cases only due to inter-surface interactions.
The filling factor contribution from an individual surface when the $i=N$ Landau level is half filled is 
$\nu_{t,b}=N$, so the partial filling factors listed in the legend of Fig.~\ref{fig:Nu024Gaps} also
specify the character of the partially occupied Landau level.  For the special case of total filling factor $\nu_{T}=0$ ($\nu_{t}=-\nu_{b}$) intersurface hybridization contributes to the gap \cite{PhysRevB.83.195413}, but 
its size vanishes quickly with thickness becoming relatively unimportant for 
thicknesses exceeding around $3 {\rm nm}$.  

In Fig.~\ref{fig:HFLLsvsV} we illustrate the dependence of the full Landau-level spectrum, including Landau level mixing, on the 
potential difference between layers for a $4{\rm nm}$ film thickness at both $\nu_{T}=0$ and $\nu_{T}=2$.  Note that the $\nu_t,\nu_b=(n,-n)$ gaps are due to spontaneous interlayer coherence and 
decrease slowly in size with increasing $n$.  This slow decrease can be traced to
partial cancellation between a decrease in exchange integral magnitude
with increasing $N$ because of form factor effects, and an increase in 
supportive Landau-level mixing effects which are enhanced by decreasing Landau level spacing.   
Although the dependence illustrated 
here is {\em vs.} Landau level index tuned by potential $V$ at fixed field, a similar robustness in the $n$-dependence
can be expected when indices are increased at fixed $V$ by reducing the magnetic field strength.
In the limit of zero field the inter-layer coherence effect illustrated here reduces to 
zero-field bilayer exciton condensation \cite{PhysRevB.77.041407, PhysRevLett.103.066402, chomoore}.  For $\nu_{T}=2$ ($\nu_{t} \ne \nu_{b}$) the gaps tend to be smaller and mean-field theory predicts Ising-like order in which charge is shifted between layers
rather than inter-layer coherence.  Ising like order is expected \cite{PhysRevB.63.035305} because of the difference 
between the form factors of the crossing levels.  For Ising order the inter-layer potential acts like an
external field which couples to the order parameter.  The jumps in the Landau level spectrum 
in the $\nu_{T}=2 $ panel are associated with pseudospin reversal near Landau level crossing points and 
reflect the hysteretic nature of the mean-field solutions in this case.  (The spectra that are shown
are for one $V$ sweep direction.) 

\noindent
{\em Discussion}---There is at present great interest in the electronic properties of TI thin films with 
Fermi levels close to their surface state Dirac points.  When electron-electron interactions are neglected 
and there is no electric potential drop across its bulk, a TI thin film has only odd integer 
quantum Hall plateaus originating from the half-quantized Hall effects on both 
surfaces.  In this Letter we have demonstrated that even integer quantum Hall plateaus are also 
expected due to electron-electron interaction effects which lead to the formation of 
quantum Hall superfluid states.  We know from experience with semiconductor quantum well bilayers that 
spontaneous coherence states are easily destroyed by disorder, particularly disorder which allows
Landau level mixing, so these states should be most easily realized when the 
$N=0$ Dirac Landau levels (which are widely separated from $N \ne 0$ levels)
on both top and bottom surfaces are half-filled, {\em i.e.} at total Landau level filling factor $\nu_T=0$.

The nature of the charged excitations
that contribute to low-temperature transport in spontaneous coherence 
states depends on the strength of interlayer tunneling \cite{PhysRevB.54.11644}.
For TI thin films, tunneling is important only for $\nu_{T}=0$ because of single-particle physics. When tunneling is not important ($\nu_{T} \ne 0$, or sample thicknesses exceeding about $3\rm{nm}$),
the charged excitations are topological meron-antimeron pairs
with layer polarization near their cores. In this regime we predict, owing to the larger values of $\rho$ in TI's compared to quantum wells, that the Kosterlitz-Thouless phase transition
will take place at higher temperatures than would be expected for quantum wells.  As in the semiconductor bilayer case, 
disorder is expected to induce merons in the bilayer ground state, complicating all coherent state 
properties.  In thinner TI's, the meron-antimeron pairs at $\nu_{T}=0$ are 
transformed by tunneling into domain wall psuedospin textures
and the sensitivity to disorder is reduced.
 
The transport phenomenology of $\nu_{T}=0$ quantum Hall systems, which do not occur in 
semiconductor bilayers, is like that of magnetic field 
$B=0$ bilayer exciton condensates  \cite{su:2008} because
the system  no longer supports non-dissipative chiral edge conduction channels.  The presence of a $\nu=0$ 
gap is readily revealed by low-temperature insulating behavior in standard transport measurements 
that contact the two surfaces independently to measure drag and 
counterflow current resistances.  It seems likely that the experimental separate-contacting  
challenge could be met, and superfluid properties studied,
if standard transport measurements revealed the formation of a 
$\nu=0$ gap. 

One important property of quantum Hall superfluids
is that they tolerate potential differences between 2D surfaces 
which shift the condensate charge more to one layer than to the other.  Spontaneous coherence 
and the associated condensation energy can be maintained while satisfying electrostatic requirements 
on charge density distributions.  This property might explain some of the quantum Hall effects 
already observed in strained HgTe systems \cite{PhysRevLett.106.126803},
which exhibit quantum Hall effects that are difficult to understand 
on the basis of independent behavior on isolated top and bottom TI surfaces.
Strained HgTe is a three-dimensional TI with a small gap and 
should have independent Dirac-like 2D electron gases (2DEG) located at opposite sides of wide quantum wells. 
Because the quantum well thickness in these experiments exceeded the magnetic 
length by a factor of $\sim 10$, the surprising quantum Hall effects 
discovered in these systems appear at first to occur at 2DEG separations that exceed
the coherence limit.  The TI surface states in this 
material are however spatially extended and could be peaked quite far from the 
wide quantum well edges, reducing the effective TI film thickness.  In addition
the same dielectric screening effects discussed here which increase the capacitance without 
substantially decreasing the charge gaps, also increase the layer separation at which coherence survives. 
For Bi$_2$X$_3$ TI's (X=Se, Te) we have estimated that the maximum $d/\ell$ ratio is increased by half compared to the
GaAs case \cite{PhysRevLett.100.106803, PhysRevLett.104.016801}. 
Future work which accounts quantitatively for the orbital character of the topologically
protected surface states in strained HgTe wide well TI's will be needed to explore the role of 
inter-surface coherence in this TI more fully.  

DT and BL were supported by One-Time Research and Research Enhancement  
Program grants, respectively, from TSU. BL also  
acknowledges computational support from High Performance Computing  
Group at TSU. AHM was supported by the Welch Foundation under Grant No. TBF1473, by NRISWAN, and by DOE Division of Materials Sciences and Engineering Grant No. DEFG03-02ER45958. 
EMH was supported by DFG grant HA5893/1-1.

\bibliographystyle{apsrev}
\bibliography{TI_sp_cohr_references}

\end{document}